\documentclass[aps,pra,superscriptaddress,twocolumn]{revtex4-1} 

\usepackage{amsmath}
\usepackage{epstopdf}
\usepackage{graphicx}
\usepackage{bm,color,bbm}
\usepackage{hyperref, mathtools}
\usepackage{caption}
\usepackage{subcaption}

\newcommand{\beq}{\begin{equation}}
\newcommand{\eeq}{\end{equation}}
\newcommand{\bqa}{\begin{eqnarray}}
\newcommand{\eqa}{\end{eqnarray}}

\newcommand{\id}{\mathbbm{1}}

\begin{document}

\preprint{}

\title{EPR Steering Inequalities from Entropic Uncertainty Relations}


\author{James Schneeloch}
\affiliation{Department of Physics and Astronomy, University of Rochester, Rochester, NY 14627}

\author{Curtis J. Broadbent}
\affiliation{Department of Physics and Astronomy, University of Rochester, Rochester, NY 14627}
\affiliation{Rochester Theory Center, University of Rochester, Rochester, NY 14627}
\author{Stephen P. Walborn}
\affiliation{Instituto de F\'{\i}sica, Universidade Federal do Rio de Janeiro, Caixa Postal 68528, Rio de Janeiro, RJ 21941-972, Brazil}
\author{Eric G. Cavalcanti}
\affiliation{School of Physics, University of Sydney, NSW 2006, Australia}
\affiliation{Quantum Group, Department of Computer Science, University of Oxford, OX1 3QD, United Kingdom}
\author{John C.  Howell}
\affiliation{Department of Physics and Astronomy, University of Rochester, Rochester, NY 14627}



\date{\today}

\begin{abstract}
We use entropic uncertainty relations to formulate inequalities that witness Einstein-Podolsky-Rosen $($EPR$)$ steering correlations in diverse quantum systems. We then use these inequalities to formulate symmetric EPR-steering inequalities using the mutual information. We explore the differing natures of the correlations captured by one-way and symmetric steering inequalities, and examine the possibility of exclusive one-way steerability in two-qubit states. Furthermore, we show that steering inequalities can be extended to generalized positive operator valued measures $($POVMs$)$, and we also derive hybrid-steering inequalities between alternate degrees of freedom.
\end{abstract}


\maketitle


\section{Introduction}
The ability to witness explicitly quantum correlations $($i.e. entanglement$)$ between arbitrary observables without having to characterize the density operator is extremely useful, and has received much attention \cite{Walborn2011,Walborn2009,Duan2000,Wiseman2007,Howell2004,Starling2011,Reid1989,Reid2009,Cavalcanti2009,Edgar2012,Handchen2012}. Entropic witnesses of entanglement, formed from the building blocks of information theory, may play an important role in the development and implementation of superior quantum information protocols such as quantum key distribution $($QKD$)$ \cite{bb84paper}. For certain tasks, such as verifying security in QKD with as few assumptions as possible, it is not sufficient only to witness entanglement \cite{Acin2007}. Fortunately, there are witnesses which detect stronger levels of quantum correlation $($e.g. Bell-nonlocality$)$ in exchange for witnessing entanglement in fewer states. Between Bell-nonlocality \cite{CHSHbell1969} and mere nonseparability \cite{Werner1989}, there is another category of entanglement known as EPR steering \cite{Wiseman2007} corresponding to a level of quantum correlation strong enough to demonstrate the EPR paradox \cite{EPR1935}, but not strong enough to rule out all models of local hidden variables $($LHV$)$. 

In this article, we develop new EPR-steering inequalities for any set of observables which share a nontrivial entropic uncertainty relation. We use those inequalities relating discrete observables to create symmetric steering inequalities based on the mutual information \cite{Cover2006}; we examine the qualitative differences in states violating one-way vs symmetric steering inequalities; we derive steering inequalities between disparate degrees of freedom useful in studying hybrid-entangled states \cite{Żukowski1991,Ma2009,Neves2009}; and we will explore applications of these steering inequalities beyond their direct use as entanglement witnesses.

\section{Foundations and Motivation}
EPR steering is the ability to nonlocally influence the set of possible quantum states of a given quantum system through the measurements on a second distant system sufficiently entangled with the first one. By choosing which observable to measure on the second system, one can ``steer'' the first system to be well-defined in any of its observables without directly interacting with it. However, one cannot know or determine in advance what the outcome of a measurement will be, as these outcomes are intrinsically random. It is only when measurement outcomes between systems are compared that we are able to see the effect of measuring one system on the other. It is this nonlocal influence that is embodied in EPR steering. It is this randomness in measurement outcomes that reinforces the no-signalling theorem \cite{nosignallingpaper} $($i.e. that rules out EPR steering as a possible means of faster than light communication$)$.

Strong correlations across conjugate observables $($e.g. in both position and momentum$)$ is a signature of entanglement, and it is these correlations that make EPR steering possible. In the original EPR situation \cite{EPR1935}, if we assume that the effect of measurement cannot travel faster than light, then any details of the observables of system $B$ obtained from measurements on system $A$ must be embedded in the local state of $B$, independent of any measurement performed on $A$. Following EPR, we ascribe inferred, ``elements of reality'', to each of these inferred properties of $B$. The paradox arises when $A$ and $B$ are so entangled that the inferred elements of reality of say, position and momentum, of $B$ are so well localized that they begin to violate uncertainty relations for single systems. If the inferred elements of reality of $B$ violate an uncertainty relation, then there cannot be a local quantum state for $B$ that reproduces such measurement results. If the inferred elements of reality of $B$ rule out a local quantum state for $B$, then this implies that it cannot both be the case that quantum correlations are local, and that conjugate observables of a given system always satisfy an uncertainty relation $($as quantum theory stipulates$)$. Unwilling to discard locality, EPR concluded that quantum mechanics must give an incomplete description of $B$.

Schr{\"o}dinger \cite{schrodinger1936} was the first to use the term ``steering'' in response to the original EPR paradox \cite{EPR1935} as a generalization beyond position and momentum. It wasn't until recently, however, that Wiseman \emph{et~al.}~\cite{Wiseman2007} formalized EPR steering in terms of the violation of a \emph{local hidden state} $($LHS$)$ model, a general class of models where, say, system $B$ has a local quantum state classically correlated with arbitrary variables at $A$. An entangled pair of systems is said to be one-way steerable if only one subsystem does not admit an LHS model. If neither subsystem admits an LHS model, the entangled pair is said to be two-way or symmetrically steerable. If $B$ has a local quantum state classically correlated with $A$, then the measurement probabilities of system $B$ must not violate any single-system uncertainty relation, even when they are conditioned on the outcomes of $A$ $($or on anything else$)$. Because of this, EPR steering is witnessed whenever conditional measurement probabilities violate an uncertainty relation. EPR steering requires entanglement because probability distributions on separable states can always be represented by an LHS model.

Though the concept of EPR steering was first formalized by Wiseman \emph{et~al.}~\cite{Wiseman2007}, Reid \cite{Reid1989} was the first to develop an experimental criterion for the EPR paradox using conditional variances and the Heisenberg uncertainty relation. A general theory of EPR-steering inequalities based on the assumption of an LHS model was developed in \cite{Cavalcanti2009}, where Reid's criterion was shown to emerge as a special case. Later, Walborn \emph{et~al.}~\cite{Walborn2011} formulated a steering inequality based on Bialynicki-Birula and Mycielski's entropic position-momentum uncertainty relation \cite{BiałynickiBirula1975}. Since their entropic uncertainty relation implies Heisenberg's uncertainty relation, the set of states witnessed by Walborn \emph{et~al.}'s steering inequality contains all the states witnessed by Reid's inequality, making Walborn \emph{et~al.}'s steering inequality more inclusive. The same is not true in the discrete case, as we will show later.

An interesting open question regarding EPR steering was raised by Wiseman \emph{et~al.} \cite{Wiseman2007}: are there states which allow steering in only one direction (say, from Alice to Bob), but not vice-versa? Some evidence that this may be the case was given for continuous-variable systems by Midgley \emph{et~al.} \cite{Midgley2010}, who showed, at least in the case where Alice and Bob are restricted to Gaussian measurements, that there are states that demonstrate steering in one direction only. Though a proof of the existence of exclusively one-way steerable states is beyond the scope  of this paper, in Sec.~6, we do extend the results of Midgley \emph{et~al.}, i.e., we show that at  least in the case considering mutually unbiased measurements, there are states which can  demonstrate steering using our inequalities in one way, but not in the other.

\section{Local Hidden State Models}
In order to develop our new entropic steering inequalities for pairs of arbitrary observables, we use the work of Walborn \emph{et~al.}~\cite{Walborn2011}, which considered the case for continuous observables as follows. Let $\hat{x}^{A}$ and $\hat{k}^{A}$ be continuous observables of system $A$ with possible outcomes $\{x^{A}\}$ and $\{k^{A}\}$, and let $\hat{x}^{B}$ and $\hat{k}^{B}$ be the corresponding observables of system $B$. According to its definition in \cite{Wiseman2007}, EPR steering occurs when the observed correlations do not admit an LHS model. The system is said to admit an LHS model if and only if the joint measurement probability density can be expressed as follows:

\begin{equation}
\rho(x^{A},x^{B}) = \int d\lambda \; \rho(\lambda) \rho(x^{A}|\lambda) \rho_{Q}(x^{B}|\lambda),
\end{equation}
where $\rho_{Q}(x^{B}|\lambda)$ is the probability density of measuring $\hat{x}^{B}$ to be $x^{B}$ given the details of preparation in the hidden variable $\lambda$. The subscript $Q$ denotes the fact that this is a probability density arising from a single quantum state, i.e. that it is a probability density arising from quantum system $B$ whose details of preparation are governed only by the hidden variable $\lambda$. On the other hand, no assumptions have been made about the origin of $A$'s probability distribution.

Using the positivity of the continuous relative entropy \cite{Cover2006} between any pair of probability distributions or densities, Walborn \emph{et~al.}~showed that it is always the case for continuous observables in states admitting LHS models that $($since the relative entropy between $\rho(x^{B},\lambda|x^{A})$ and $\rho(\lambda|x^{A})\rho(x^{B}|x^{A})$ is always greater than zero,$)$
\begin{equation}\label{CVSE}
h(x^{B}|x^{A})\geq \int d\lambda \; \rho(\lambda) h_{Q}(x^{B}|\lambda),
\end{equation}
where $h_{Q}(x^{B}|\lambda)$ is the continuous Shannon entropy arising from the probability density $\rho_{Q}(x^{B}|\lambda)$. 

In developing our steering inequalities for arbitrary observables, we note that the same arguments used to develop LHS constraints for continuous observables can be used to formulate LHS constraints for discrete observables as well. Consider discrete observables $\hat{R}^{A}$ and $\hat{S}^{A}$ with outcomes $\{R^{A}_{i}\}$ and $\{S^{A}_{i}\}$, respectively, and where $i$ runs from $1$ to the to the total number of distinct eigenstates $N$.  Let $\hat{R}^{B}$ and $\hat{S}^{B}$ be the corresponding observables for system $B$.  Since the positivity of the relative entropy is a fact for both continuous and discrete variables, we can derive the corresponding LHS constraint for discrete observables in the same way;
\begin{equation}\label{discSE}
H(R^{B}|R^{A})\geq \sum_{\lambda} P(\lambda) H_{Q}(R^{B}|\lambda),
\end{equation}
where $H_{Q}(R^{B}|\lambda)$ is the discrete Shannon entropy of the probability distribution $P_{Q}(R^{B}|\lambda )$, where again the subscript, ``$Q$'', means that it corresponds to a quantum state. All observables of systems admitting LHS models must obey inequality \eqref{discSE} for discrete observables, or \eqref{CVSE} for continuous observables.

\section{Entropic Steering Inequalities}
Consider the right hand side of inequality \eqref{CVSE}. Where position $\hat{x}$ and wavenumber $\hat{k}$ are continuous observables constrained by the entropic uncertainty relation \cite{BiałynickiBirula1975}
\begin{equation}\label{eupcont}
h_{Q}(x^{B}) + h_{Q}(k^{B})\geq \log(\pi e),
\end{equation}
we readily see that if we take a weighted average of these entropies with weight function $\rho(\lambda)$, we get the right hand side of \eqref{CVSE}. From there, it is straightforward to show $($as Walborn \emph{et~al.} did$)$, that any state admitting an LHS model in position-momentum must satisfy the inequality,
\begin{equation}\label{stineqcont}
h(x^{B}|x^{A}) + h(k^{B}|k^{A}) \geq \log(\pi e).
\end{equation}
Indeed, for any pair of continuous observables with an entropic uncertainty relation resembling \eqref{eupcont}, there is always the corresponding steering inequality \eqref{stineqcont}. 

As explained in Ref.~\cite{Maassen1988}, given any pair of discrete observables $\hat{R}$ and $\hat{S}$ in the same $N$-dimensional Hilbert space, with eigenbases $\{|R_{i}\rangle\}$ and $\{|S_{j}\rangle\}$, respectively $($such as for different components of the angular momentum$)$, there exists the entropic uncertainty relation
\begin{align}\label{Maassenineq}
&H_{Q}(R) + H_{Q}(S) \geq\log(\Omega)\\ \label{bb}
&:\:\Omega \equiv \min_{i,j} \bigg(\frac{1}{|\langle R_{i}|S_{j}\rangle|^{2}}\bigg).
\end{align}
When the discrete observables $\hat{R}$ and $\hat{S}$ are maximally uncertain with respect to one another, all measurement outcomes of one observable are equally likely when the system is prepared in an eigenstate of the other observable. These maximally uncertain observables $($termed mutually unbiased$)$ have an uncertainty relation where $\Omega$ obtains its maximum value given by the dimension $N$ of the Hilbert space. The uncertainty relation is saturated when the system is prepared in an eigenstate of one of the unbiased observables.

Using the discrete entropic uncertainty relation \eqref{Maassenineq} along with our LHS constraint for discrete observables \eqref{discSE}, we immediately arrive at a new entropic steering inequality for pairs of discrete observables
\begin{equation}\label{discWSE}
H(R^{B}|R^{A}) + H(S^{B}|S^{A}) \geq\log(\Omega^{B}).
\end{equation}
where $\Omega^{B}$ is the value $\Omega$, given in definition\eqref{bb} associated with the observables $\hat{R^{B}}$ and $\hat{S^{B}}$.
 
For quantum systems in which conjugate bases are discrete and continuous, such as with angular position and angular momentum, the entropic uncertainty relation will have a sum of both discrete and continuous entropies. This doesn't give rise to any complications because the LHS constraints deal with only one measured observable at a time.
Given a continuous observable $\hat{x}$ and a discrete observable $\hat{R}$ with uncertainty relation \cite{BialynickiBirula1985},
\begin{equation}\label{inbetweenineq}
h_{Q}(x) + H_{Q}(R) \geq C,
\end{equation}
where $C$ is a real-valued placeholder dependent on the particular uncertainty relation, we readily find a new steering inequality between a discrete and a continuous observable,
\begin{equation}
h(x^{B}|x^{A}) + H(R^{B}|R^{A}) \geq C.
\end{equation}
In fact, an EPR-steering inequality of this type has recently been experimentally tested for discrete and continuous components of position and momentum variables of entangled photons \cite{leach2010,carvalho12}.

\section{Symmetric Steering Inequalities}
Up until now, all the EPR-steering inequalities discussed here have been asymmetric between parties; they rely on conditional probability distributions, and their violation rules out LHS models from describing only one of the parties' measurements. Violating a more restrictive EPR-steering inequality that is symmetric between parties would allow one to rule out LHS models for both parties at the same time.

Cavalcanti \emph{et~al.} \cite{Cavalcanti2009} were the first to develop such a symmetric steering inequality by showing that the variance of sums and differences always exceeds the largest of the conditional variances in Reid's inequality \cite{Reid1989}. For position and momentum, the sum/difference steering inequality takes the form
\begin{equation}
\sigma^{2}(x^{A} \pm x^{B}) \sigma^{2}(k^{A} \mp k^{B}) \geq \frac{1}{4},
\end{equation}
which is just Mancini \emph{et~al.}'s separability inequality \cite{Mancini2002} with a tighter bound of $\frac{1}{4}$ instead of $1$.

It turns out that we can also create an entropic steering inequality using sums and differences for the same reason as we now show. The entropy of a sum or difference of two random variables is never less than the larger of the two conditional entropies.
\begin{align}
h(x^{A}\pm x^{B})&\geq \max \{h(x^{A}\pm x^{B}|x^{A}),h(x^{A}\pm x^{B}|x^{B})\}\nonumber\\
&= \max \{h(x^{A}|x^{B}),h(x^{B}|x^{A})\}
\end{align}
This is true for both discrete and continuous random variables, which allows us to assert that both
\begin{equation}\label{pmsteercv}
h(x^{A}\pm x^{B}) + h(k^{A}\mp k^{B}) \geq \log(\pi e)
\end{equation}
and
\begin{equation}\label{pmsteerdv}
H(R^{A}\pm R^{B}) + H(S^{A}\mp S^{B}) \geq \log(\Omega)
\end{equation}
are valid steering inequalities coming from \eqref{stineqcont}, and\eqref{discWSE} respectively, but which are symmetric between parties, and witness EPR steering both ways at the same time. We note also that inequality \eqref{pmsteercv} is just Walborn \emph{et~al.}'s 2009 separability inequality \cite{Walborn2009} with the tighter bound $\log(\pi e)$ instead of $\log(2 \pi e)$. Whether the symmetric steering inequality \eqref{pmsteerdv} is similarly a separability inequality with a tighter bound is the subject of ongoing investigation.

These new symmetric steering inequalities \eqref{pmsteercv} have the added benefit of not needing to measure full joint probability distributions, let alone reconstructing density operators to witness that a state is EPR-steerable. The functions $x^{A}\pm x^{B}$ and $k^{A}\mp k^{B}$ as well as their discrete counterparts are commuting observables that can be measured directly in many physical systems, which means in those cases where these inequalities can be violated, it takes fewer measurements to witness that a state is EPR-steerable.

However, there's a subtle but important point to be noted here. Demonstration of EPR steering through these sum/difference inequalities requires that the observables $x^{A}$, $x^{B}$, $k^{A}$, and $k^{B}$ be measured individually; violation of these inequalities through a direct measurement of $x^{A}\pm x^{B}$ does not, strictly speaking, demonstrate EPR-steering (or equivalently, demonstrate the EPR paradox). This is because: $($i$)$ to determine which experimental procedure corresponds to $x^{A}\pm x^{B}$, etc. requires extra assumptions about the quantum operators corresponding to Alice's measurements which goes beyond the assumption of an LHS model; and $($ii$)$ measurements of the sum/difference observables require that we interact the systems $A$ and $B$, undermining the assumption of locality. On the other hand, violation of these  sum/difference inequalities does imply that the state is EPR-steerable in the sense that \emph{if} the individual measurements were performed instead, those statistics would not be describable by an LHS model. This might be useful when the objective of the experiment is to characterize the state rather than a fundamental demonstration of nonlocality.

A useful property of discrete observables and discrete approximations to continuous ones \cite{BialynickiBirula1984} is that the Shannon entropies are bounded above either by the logarithm of the dimensionality of the system $N$, or by the number of discrete windows into which the observable is partitioned. With this upper bound, we can create symmetric EPR-steering inequalities using the mutual information.

The mutual information of the joint probability distribution of measurement outcomes of $\hat{R}^{A}$ and $\hat{R}^{B}$ is defined as
\begin{align}
I(R^{A}:R^{B}) &\equiv H(R^{A}) + H(R^{B}) - H(R^{A},R^{B})\\
&=H(R^{B}) - H(R^{B}|R^{A}).\nonumber
\end{align}
We can express the steering inequality \eqref{discWSE} in terms of the mutual information and use the maximum possible values of the marginal entropies to arrive at a general symmetric steering inequality;
\begin{equation}\label{mutinfineq}
I(R^{A}:R^{B}) + I(S^{A}:S^{B}) \leq \log\bigg(\frac{N^{2}}{\min \{\Omega^{A},\Omega^{B}\}}\bigg).
\end{equation}
This mutual information inequality yields some important insights. We choose the minimum of $\{\Omega^{A},\Omega^{B}\}$ since we want this symmetric steering inequality to witness steering both ways, i.e. to rule out LHS models for both parties.

Consider the case where $\hat{R}^{A}$ and $\hat{S}^{A}$ $($and similarly $\hat{R}^{B}$ and $\hat{S}^{B}$$)$ are mutually unbiased observables. Their uncertainty relation reaches the maximum lower bound, where $\Omega^{A}=\Omega^{B}=N$, which makes the bound on the right hand side of \eqref{mutinfineq} $\log(N)$. This maximal bound is also equal to the largest possible value of the mutual information $I(R^{A}:R^{B})$ or $I(S^{A}:S^{B})$. If $\hat{R}$ and $\hat{S}$ $($for either $A$ or $B$$)$ were somewhere between being mutually unbiased and simultaneously measurable, the mutual information bound would be between $\log(N)$ and $2\log(N)$; at the upper limit, the observables commute. 

Though a pair of quantum systems can be classically prepared $($i.e. with local operations and classical communication$)$ to be strongly correlated in one variable, quantum entanglement is required to have strong simultaneous correlations in observables which are mutually unbiased, that is, strong enough to violate an EPR-steering inequality. Indeed, if a pair of systems were perfectly correlated in one observable, any correlation in a conjugate observable is sufficient to demonstrate symmetric EPR steering in particular and entanglement in general. 

Conditional and symmetric steering inequalities witness different levels of nonlocality. While violating a conditional steering inequality rules out an LHS model for either party $A$ or $B$, violating a symmetric steering inequality rules out LHS models for both parties $A$ and $B$. It is important to know whether these steering inequalities witness entanglement in qualitatively different sets of states, or if their violation is merely a signpost of progressively stronger entanglement. To answer this question, we must determine what differences there are in the sets of states that violate each inequality.

Let $V_{C}$ be the difference between the bound and sum of conditional entropies in the discrete conditional steering inequality on party B \eqref{discWSE} $($i.e. the violation of \eqref{discWSE} in number of bits$)$, and let $V_{M}$ be the difference between the sum of mutual informations and the bound in the discrete symmetric steering inequality \eqref{mutinfineq}. Here, $V_{C}$ and $V_{M}$ are positive for positive violation and we limit ourselves for simplicity to observables where $\Omega^{A}=\Omega^{B}\equiv\Omega$. The difference, $V_{C}-V_{M}$, is expressed as
\begin{equation}
V_{C} - V_{M} = 2\log(N) - \big(H(R^{B}) + H(S^{B})\big).
\end{equation}
From this we know immediately that the violations are the same, $V_{C}=V_{M}$ if and only if the marginal  measurement probability distributions are both uniform $($e.g. that the density operator is one whose marginal states are maximally mixed$)$.

\begin{figure*}[t]
 \centering
\includegraphics[width=\textwidth]{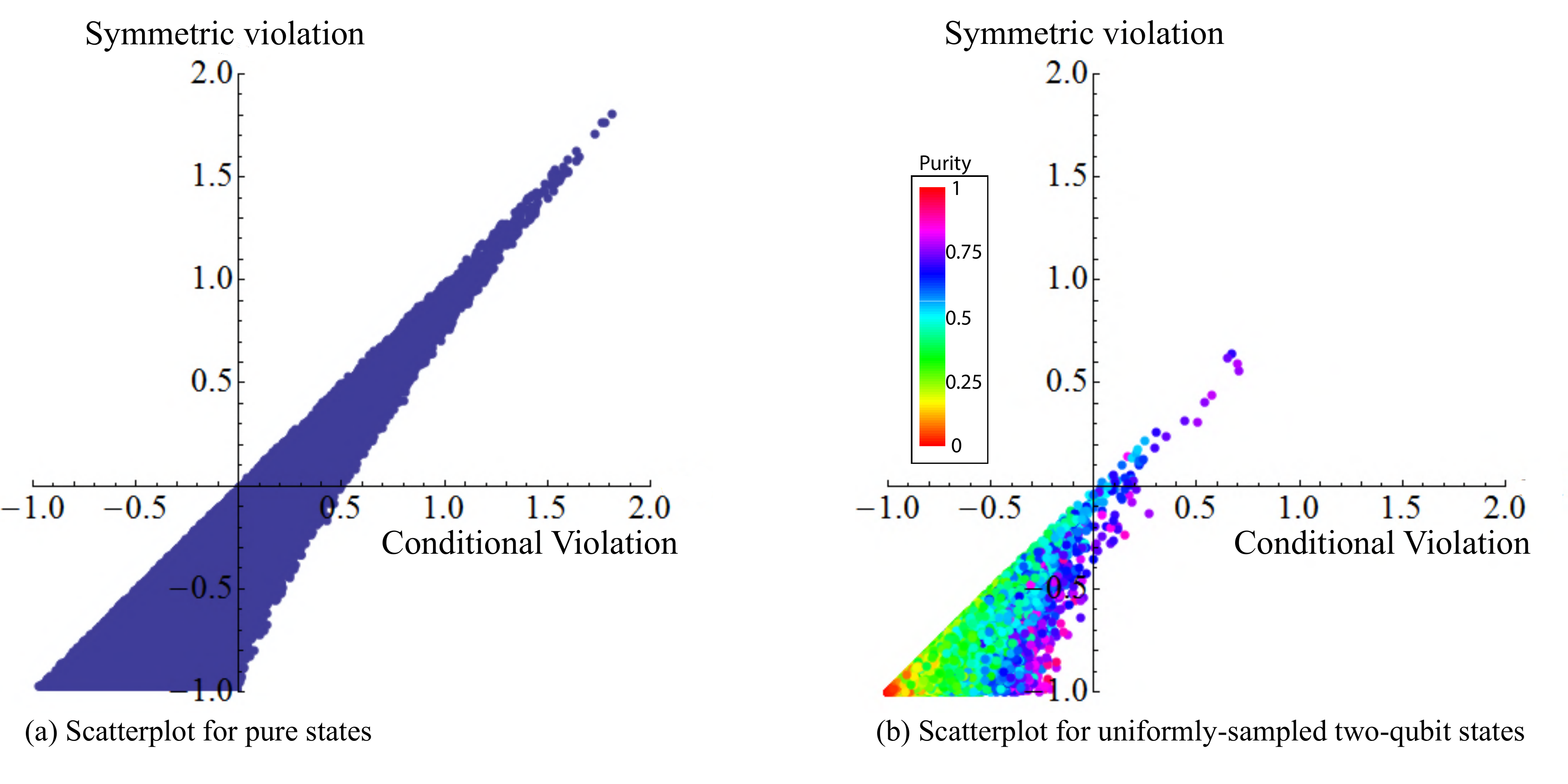}
\caption{Scatterplots of the violation of the conditional and symmetric steering inequalities which use all mutually unbiased bases. Each point is a random 2-qubit state. Fig.~1b is color coded according to purity, $P$, as measured by the von Neumann entropy, scaled and inverted so that 0 is maximally mixed and 1 is pure: $P = 1-\frac{S(\rho)}{2}$. The well-defined diagonal line through the origin indicates that no matter the orientation of the mutually unbiased bases, the symmetric violation never exceeds the conditional violation. The plots thin out to the upper right since maximally entangled states are rare when uniformly sampling over pure states, and rarer still when uniformly sampling over all states.}
\end{figure*}

Since the Shannon entropies $H(\hat{R}^{B})$ and $H(\hat{S}^{B})$ are bounded below by the underlying von Neumann entropy $S(\hat{\rho}^{B})$ \cite{nielsen2000}, which is in turn bounded below by the entanglement of formation $E(\hat{\rho})$ \cite{nielsen2000}, we see that the largest possible difference in violations decreases with increasing entanglement:
\begin{equation}
V_{C} - V_{M} \leq 2\big(\log(N) - S(\hat{\rho}^{B})\big) \leq 2\big(\log(N) - E(\hat{\rho})\big).
\end{equation}
This agrees with our previous result since maximally entangled states also have maximally mixed marginal probability distributions. Indeed, since the largest possible value for the violations is the same in both inequalities, we expect there to be no difference in violations for maximally entangled states. This is particularly well illustrated in Figures ~1a and 1b where we simulated random 2-qubit states to compare $V_{C}$ and $V_{M}$ for the conditional and symmetric steering inequalities (\ref{eq:MUB_cond},~\ref{mutinfineq2}) using all mutually unbiased observables as discussed in the next section. In order to generate random two-qubit states, we use the methods discussed in Ref.~\cite{Starling2011}.

It is important to note that these inequalities are only \emph{witnesses} for steering. Because violation of these inequalities are sufficient, but not necessary conditions for EPR steering, we can have states that are symmetrically steerable, but which fail to violate both kinds of steering inequalities presented here. If a state violates a conditional steering inequality and not a symmetric steering inequality, we know it is at least one-way steerable, but it may yet violate a different symmetric steering inequality, or indeed a different one-way steering inequality in the other direction.

One is tempted to think that because all EPR-steerable states form a proper subset of all entangled states $($and a proper superset of all Bell-nonlocal states$)$, there might be some finite nonzero threshold to the entanglement needed in a state to demonstrate EPR steering. In fact, it turns out that at least pure states with very little entanglement can, in principle, demonstrate EPR steering. This was effectively proven \cite{Yu2012} by generalizing Gisin's theorem \cite{Gisin1991} for any pair of discrete quantum systems, which states that any pure bipartite state that isn't a product state is Bell-nonlocal $($and so also EPR-steerable$)$, even for very small entropies of entanglement. A proof of Gisin's theorem for continuous variables remains an open topic for investigation.

\section{EPR Steering using all unbiased observables}
Up to this point, the discussion has been limited to uncertainty relations between pairs of observables. We must remember that for any entropic uncertainty relation, even those relating more than two observables, there is a corresponding EPR-steering inequality. Sanchez-Ruiz \cite{SánchezRuiz1995} developed entropic uncertainty relations for complete sets of pairwise complementary $($mutually unbiased$)$ observables $\{\hat{R}_{i}\}$, where $i=\{1,...,N\}$. When $N$, the dimensionality of the system, is a positive integer power of a prime number, it has been shown \cite{Durt2010} that there are complete sets of $N+1$ mutually unbiased observables.

When $N$ is even, we have the uncertainty relation
\begin{equation}
\sum_{i=1}^{N+1} H(R_{i})\geq \frac{N}{2}\log\bigg(\frac{N}{2}\bigg) +\bigg(\frac{N}{2}+1\bigg)\log\bigg(\frac{N}{2}+1\bigg)\equiv G_{even}
\end{equation}
and for odd $N$, we have
\begin{equation}
\sum_{i=1}^{N+1} H(R_{i})\geq (N+1)\log\bigg(\frac{N+1}{2}\bigg)\equiv G_{odd}.
\end{equation}
Here, $G_{even}$ and $G_{odd}$ are defined as the bounds for these uncertainty relations to condense these expressions later on. These uncertainty relations can be adapted into EPR-steering inequalities readily by substituting conditional entropies for marginal ones. Using knowledge of the purity of the state, the bounds can be improved \cite{SánchezRuiz1995, Berta2010}, but using such improved uncertainty relations requires information about the quantum state beyond the measured joint probabilities.

In the same way as was done to derive \eqref{discWSE}, we see that for $N$ even, we have the EPR-steering inequality
\begin{equation}
\sum_{i=1}^{N+1} H(R_{i}^{B}|R_{i}^{A}) \geq G_{even}, \label{eq:MUB_cond}
\end{equation}
and in the same way as was done to derive \eqref{mutinfineq}, we have for even $N$,
\begin{equation}\label{mutinfineq2}
\sum_{i=1}^{N+1} I(R_{i}^{A}: R_{i}^{B}) \leq (N+1)\log(N) - G_{even}.
\end{equation}
For odd $N$, we have the same expressions (\ref{eq:MUB_cond},\ref{mutinfineq2}) with $G_{odd}$ substituted in for $G_{even}$. 

As a particular example, consider the case of a pair of qubits. $N = 2$, which makes $G_{even} = 2$. The full symmetric steering inequality for a pair of qubits becomes
\begin{equation}
\sum_{i=1}^{3} I(R_{i}^{A}: R_{i}^{B}) \leq 1.
\end{equation}
which not only proves the entanglement witness first postulated in \citet{Starling2011}, but also shows that it is a symmetric steering inequality whose violation demonstrates the EPR paradox.

We note that while similar EPR-steering inequalities exist for measuring the strength of linear correlations \cite{Saunders2010}, they don't register the same information-significant behavior as inequality \eqref{mutinfineq2} for the same reason that variances don't capture as much of the necessary information about the uncertainty in a probability distribution as entropies can. Covariance and other measures of correlation are sensitive to specific functional dependence between random variables $($particularly linear dependence$)$, while the mutual information captures correlations between random variables whose dependence may be entirely arbitrary, but still well-determined.

\section{Violations of Steering inequalities by Quantum States}

For simplicity, we now look for violations of our inequalities in entangled two-qubit states. We first examine the Werner states \cite{Werner1989}, defined as
\begin{equation}
W_p = p |\Phi_s\rangle \langle \Phi_s| + (1-p) \frac{\id}{4}, \label{eq:Werner}
\end{equation}
where $|\Phi_s\rangle$ is the maximally-entangled singlet state, $\id/4$ is the maximally mixed state for two qubits, and $p$ is the weight of the singlet state in $W_p$. These states were shown in \cite{Wiseman2007} to be steerable in principle (i.e. with an infinite number of measurements) for all values of $p>1/2$. In practice, (i.e. with finite numbers of measurements), this is not achievable. In \cite{Cavalcanti2009} it was shown that these states violate a linear steering inequality with two measurement settings at each side for $p>1/\sqrt{2} \approx 0.71$ and with three measurement settings for  $p>1/\sqrt{3} \approx 0.58$, and violates a variance-based steering inequality for $p>(\sqrt{5}-1)/2\approx0.62$ and $p>1/\sqrt{3}\approx0.58$, with three and four measurement settings for Bob, respectively $($the latter inequality was introduced in \cite{Cavalcanti2009b}$)$.

We first apply the Werner state to our conditional steering inequality \eqref{discWSE}, with measurements in the Pauli $X$ and $Z$-bases on each side. The inequality then reads as,
\begin{equation}\label{biqubitsteer}
H(\sigma_{x}^{B}|\sigma_{x}^{A}) + H(\sigma_{z}^{B}|\sigma_{z}^{A}) \geq 1.
\end{equation} 
For the Werner state, the left hand side of this inequality \eqref{biqubitsteer} reduces to
\begin{align}
&H(\sigma_{x}^{B}|\sigma_{x}^{A}) + H(\sigma_{z}^{B}|\sigma_{z}^{A})= \\
&=-\{(1+p)\textrm{log}[(1+p)/2]+(1-p)\textrm{log}[(1-p)/2]\}\nonumber\\\nonumber
\end{align}
and violation occurs for all values of $p\gtrsim 0.78$. For our three-setting inequality \eqref{eq:MUB_cond}, we use measurements in the $X$, $Y$, and $Z$-bases, and thus for $N=2$, the inequality \eqref{eq:MUB_cond} reads as 
\begin{equation}
H(\sigma_{x}^{B}|\sigma_{x}^{A}) + H(\sigma_{y}^{B}|\sigma_{y}^{A})+ H(\sigma_{z}^{B}|\sigma_{z}^{A})\geq 2.
\end{equation} 
Applied to the Werner state, the left side is now $-3/2\{(1+p)\textrm{log}[(1+p)/2]+(1-p)\textrm{log}[(1-p)/2]\}$, and the inequality is violated for all $p\gtrsim 0.65$.

For states with completely mixed marginals, and when $\Omega^{B}=\Omega^{A}$, there is no difference between the violation of our symmetric inequality \eqref{mutinfineq} and our conditional steering inequality \eqref{discWSE}. This is also true for our steering inequalities using complete sets of mutually unbiased bases, \eqref{mutinfineq2} and \eqref{eq:MUB_cond}. The Werner state thus violates the symmetric inequalities \eqref{mutinfineq} and \eqref{mutinfineq2} in the same regimes as it violates the corresponding conditional inequalities, as calculated above. This is not surprising, since the Werner state is symmetric between parties.

It is somewhat surprising, however, that the violations of the entropic steering inequalities presented here occur for a smaller range of Werner states than do the variance-based inequalities in \cite{Cavalcanti2009}. This is fundamentally different from the result shown for the continuous-variable case by Walborn \emph{et~al.}. Those authors showed that the entropic steering inequality here reproduced as Eq.~\eqref{stineqcont} detects steering in a larger class of states than the variance-based Reid criterion. In the continuous variable case, Heisenberg's variance uncertainty relation is implied by Bialynicki-Birula and Mycielski's entropic uncertainty relation \eqref{eupcont}. Because of this, all states violating Reid's criterion must also violate Walborn \emph{et~al.}'s steering inequality. The same is not true for finite discrete variables since the maximum entropy state with a well defined variance is no longer a Gaussian, but a uniform distribution. Certainly for two-level discrete systems, there is not much qualitative difference in characterizing the uncertainty with entropies or with variances. For higher dimensions, however, entropic measures of uncertainty are superior because a sharply peaked bimodal distribution is much more well determined $($and so has much smaller entropy$)$ than a single-peaked distribution of the same variance.

\begin{figure*}[t]
\includegraphics[width=\textwidth]{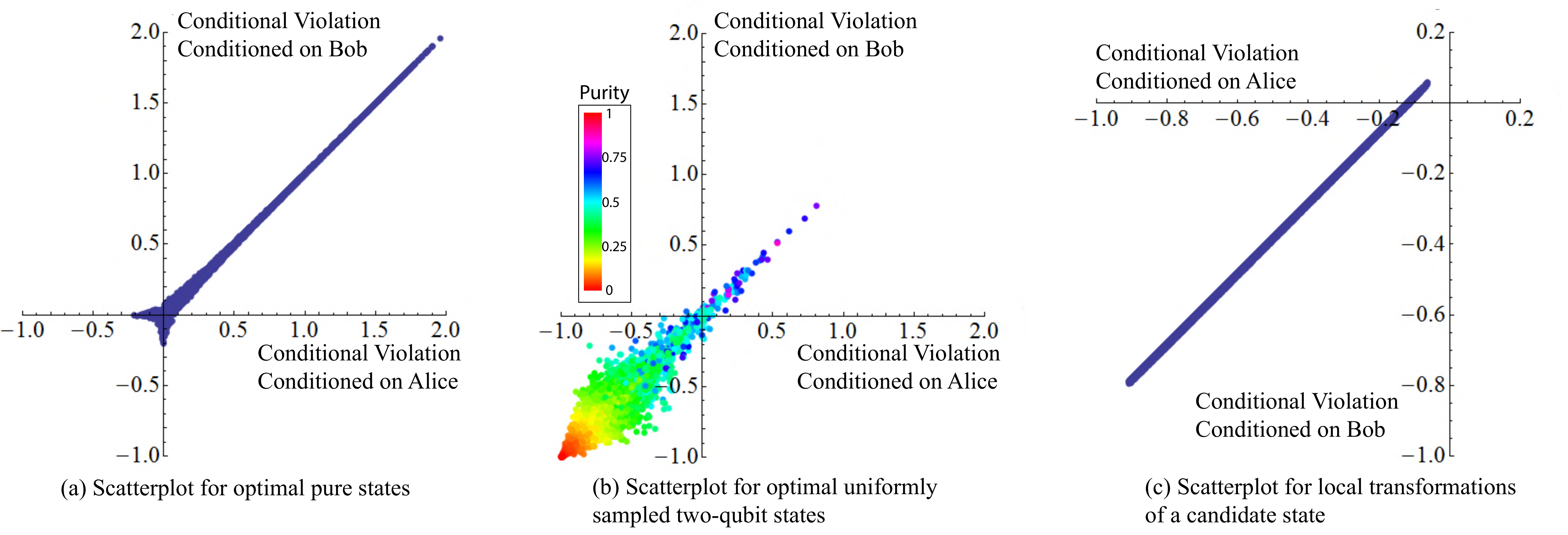}
\label{fig2}
\caption{Scatterplots of the violation in number of bits of the conditional steering inequalities when using an optimal set of mutually unbiased bases. The violation of the inequality conditioned on Alice's measurements is plotted against the violation of the inequality conditioned on Bob's measurements. In Figs.~2a and 2b, each point is a random 2-qubit state whose set of measurement bases have been selected for maximum violation. For higher violation, the scatterplots approach a diagonal line, where more entangled states tend to be more symmetric between parties.  Fig.~2c examines the violation of a candidate 2-qubit state in many different independently random sets of measurement bases. States in the upper left quadrants are ones where Alice's uncertainty is less than Bob's uncertainty. In the lower right quadrants, Bob's uncertainty is less than Alice's. Fig.~2b is color coded according to purity as defined in Fig.~1b.}
\end{figure*}

While the Werner states violate both symmetric and conditional entropic steering inequalities in the same manner, the same does not happen for all entangled states, as illustrated in Figures~1a and 1b, which survey the violation of the symmetric inequality (\ref{mutinfineq2}) vs the violation of the asymmetric $($conditional$)$ inequality (\ref{eq:MUB_cond}) for large distributions of random two-qubit states. Figure~1a examines these violations for $10^{5}$ uniformly-sampled pure states, while Figure~1b examines these violations for $10^{5}$ uniformly-sampled arbitrary states. The sampling method is described in \cite{Starling2011}. States in the lower right quadrant violate the conditional inequality, but not the symmetric inequality. The well-defined diagonal line in these plots shows that a state never violates the symmetric steering inequality \eqref{mutinfineq2} by a larger amount than the conditional steering inequality \eqref{eq:MUB_cond}, as expected.

To further demonstrate the asymmetry between parties, we surveyed the violation of the conditional inequality (\ref{eq:MUB_cond}) in the Alice-Bob direction versus the violation of the conditional inequality in the Bob-Alice direction $($seen in Figures~2a and 2b$)$ for a large distribution of states whose set of measurement bases has been chosen to maximize violation in both directions. In Fig.~2a, each point is one of $5\times10^{3}$ pure two-qubit states sampled uniformly, each of which is measured in 500 different sets of measurement bases, chosen randomly and independently by Alice and Bob, to find the one which maximizes violation in both directions. Figure~2b samples $5\times10^{3}$ general $($not necessarily pure$)$ two-qubit states, each one similarly optimized using 500 different sets of measurement bases. 

The states in the second and fourth quadrants of Figs.~2a and 2b violate our entropic conditional steering inequality in only one direction. Note however, that these results do not imply that there are no other inequalities which could demonstrate steering in the other direction. We know a priori that no pure state is exclusively one-way steerable because pure states are fundamentally symmetric between parties. As shown by a Schmidt decomposition, the sets of eigenvalues of the reduced density operators of a pure bipartite state are identical, which means their marginal statistics must be identical as well with the right set of measurement bases. In particular, for every set of measurement bases giving a particular value for the sum of conditional entropies $H(A|B)$, there must exist another set of measurement bases giving the same value for the sum of conditional entropies $H(B|A)$.  An optimal choice of local measurement basis requires that if the pure state is steerable one way, it must be steerable the other way as well. Those points in the off-diagonal quadrants of Fig.~2a are due to our inequalities being sufficient, but not necessary criteria for EPR steering. What is not clear is whether there are mixed states that may be exclusively one-way steerable.

As seen in Fig.~2b, we find some mixed states which are candidates for being exclusively one-way steerable, that is, which may allow only one-way steering even when all possible sets of measurement bases are considered. In Fig.~2c, we plot the violations of one of these mixed states in $10^{5}$ different 
measurement bases  chosen randomly and independently by Alice and Bob to see what effect 
measurement basis has on an experimenter's ability to violate our steering inequalities. There is a striking linear trend in this plot, which suggests that the difference between violations in either direction is nearly constant, that either Alice's or Bob's advantage in witnessing EPR steering is nearly independent of their choice of measurement basis $($and  therefore fundamental to the state itself$)$. We examined this trend in over 300 arbitrary random density matrices, and it is found to a varying degree in all states observed. The same trend is also seen when Alice and Bob's measurement bases are fixed to be equal to one another, though without the extra degree of freedom, finding optimal measurement bases is less likely. The trend is more pronounced for states with higher optimal violation, and diminishes in states with lower maximal violation. Though our inequalities cannot witness exclusive one-way steerability, our studies suggest that there is a fundamental asymmetry between parties in two-qubit systems whose marginal states have different purities. Again, we must reiterate that since the violation of an EPR-steering inequality is a sufficient, but not necessary condition for the state to be EPR-steerable, what we do is rule out all but those candidate states from being exclusively one-way steerable.

\section{Steering with POVM's}
Before going further, we note that Maassen and Uffink's uncertainty relation \eqref{Maassenineq} relies on $\hat{R}$ and $\hat{S}$ having nondegenerate eigenvalues. Since then, more general entropic uncertainty relations have been discovered \cite{Krishna2002} which allow $\hat{R}$ and $\hat{S}$ to be any pair of discrete observables $($without changing the form of the uncertainty relation$)$. In addition, Krishna and Parasarathy \cite{Krishna2002} have shown that for any set of generalized measurements, i.e., any POVMs $($Positive Operator Valued Measures$)$, with measurement operators $\{F_{i}\}$ and $\{G_{j}\}$
\begin{align}\label{povmineq}
&H(F) + H(G) \geq\log(\Omega_{\text{POVM}}),\\
&:\:\Omega_{\text{POVM}} \equiv \min_{i,j} \bigg(\frac{1}{||F_{i}G_{j}||^{2}}\bigg)\\
&:\:||F||\equiv \max_{|\psi\rangle} \sqrt{|\langle\psi|F^{\dagger}F|\psi\rangle|}.
\end{align}
This uncertainty relation \eqref{povmineq} allows us to create steering inequalities for POVMs in the same way as was done for projective measurements. The LHS constraints are contingent only upon measurement probabilities adhering to entropic uncertainty relations, not on those measurements being projective. If we let $\{F^{A}_{i}\}$ and $\{G^{A}_{j}\}$ be discrete sets of POVMs on party $A$, and let $\{F^{B}_{i}\}$ and $\{G^{B}_{j}\}$ be sets of POVMs obeying entropic uncertainty relation \eqref{povmineq}, it can be readily shown that
\begin{equation}
H(F^{B}|F^{A}) + H(G^{B}|G^{A}) \geq \log(\Omega_{\text{POVM}}^{B})
\end{equation}
is a valid steering inequality for POVMs where $\Omega_{\text{POVM}}^{B}$ is $\Omega_{\text{POVM}}$ for measurements on party $B$. Since we no longer have to restrict ourselves to projective Von Neumann measurements, we can study EPR steering when we can only interact indirectly with the system as with weak measurements \cite{aharonov1988}.

\section{Hybrid Steering Inequalities}
Our steering inequalities were formed from pairs or sets of non-commuting observables on a single quantum system conditioned on the corresponding observables of another quantum system. However, the derivation of our steering inequalities does not require the observables on the second system to be the same as those of the first. For example, in the inequality derived by Walborn \emph{et~al.} \eqref{stineqcont}, we require that $x^{B}$ and $k^{B}$ be conjugate to one another in accordance with the uncertainty relation \eqref{eupcont}. The observables $x^{A}$ and $k^{A}$ need not be the position and momentum of system $A$ (respectively) to have a valid steering inequality; any pair of observables for system $A$ will do. In fact, we can even condition both $x^{B}$ and $k^{B}$ on the same observable; this would make a valid steering inequality, though it would be impossible to violate in principle because conditioning on only one observable of system $A$ only changes what would be the local state of system $B$ from which one draws measurement probabilities. In this case,  all measurements are made on the same state of system $B$, which must satisfy the uncertainty relation \eqref{eupcont}.

With this additional freedom in deriving steering inequalities, we can examine entanglement between different degrees of freedom. For example, violation of
\begin{equation}\label{crosssteerineq}
h(x_{B}|\sigma_{z A}) + h(k_{B}|\sigma_{y A}) \geq \log(\pi e)
\end{equation}
or
\begin{equation}\label{crosssteering2}
H(\sigma_{z A}|x_{B}) + H(\sigma_{y A}|k_{B}) \geq 1.
\end{equation}
witnesses EPR steering between the position-momentum degree of freedom of one system, and the spin-polarization of the other. By using the discrete uncertainty relation for coarse-grained position and momentum \cite{BialynickiBirula1984}, we can witness such entanglement in the laboratory. We call these steering inequalities between different degrees of freedom \emph{hybrid}-steering inequalities. Hybrid-steering inequalities may prove useful in the study of hybrid-entangled states, i.e., states which are entangled across different degrees of freedom \cite{Neves2009}.

\section{Steering and QKD}

In classical information theory \cite{Cover2006}, the mutual information can be interpreted as the channel capacity of a communication system with source at party $A$ and receiver at party $B$ $($or also the other way around$)$, giving our EPR-steering inequalities special utility in quantum information protocols.  In particular, security in quantum key distribution $($QKD$)$ schemes requires that Alice and Bob are able to prove that the quantum systems transmitted on quantum channels have not been intercepted by Eve. 

Recently, it has been shown that EPR steering is linked to the secret key rate in one-sided device-independent quantum key distribution $($1sDIQKD$)$ \cite{BranciardQKD2012}.  1sDIQKD lies between conventional QKD and full device-independent QKD \cite{Mayers1998,Acin2007} in that only one of the users trusts his/her measurement device.  This connection was shown for asymmetric EPR steering.  It is thus an interesting question as to what link can be made between the symmetric EPR-steering inequalities and secure transmission rates in a quantum channel.

Intuitively, violating a symmetric EPR-steering inequality rules out the possibility that Eve performs independent (incoherent) attacks on either Alice or Bob's channel all of the time, since enough of the joint states shared by Alice and Bob must be correlated enough to rule out local hidden states for both parties. Thus if Eve is constrained to perform only incoherent attacks on either party, violating a symmetric steering inequality should guarantee a nonzero secret key rate, since some of Alice's and Bob's shared systems would have to have been untouched by Eve, meaning that Eve could not have a perfect LHS model for all of Alice and Bob's systems.

In the more general situation, Eve cannot be limited to incoherent attacks, though it is still possible to formulate secret key rates in terms of the mutual information \cite{Scarani2009}.   For now it remains an open question as to whether the degree of violation of our symmetric  EPR-steering inequalities (in bits) provides a lower bound to the secure key transmission rate.

\section{Conclusion}
In this paper, we have shown how any set of operators obeying an entropic uncertainty relation can give rise to an entropic steering inequality. Specifically: we've derived steering inequalities for pairs of arbitrary observables; we've derived steering inequalities for complete sets of mutually unbiased observables; we've derived symmetric steering inequalities and hybrid-steering inequalities; and we've derived steering inequalities for POVM's. In addition, we have examined the possibility of exclusive one-way steering in two-qubit states, and looked at possible applications for these steering inequalities in QKD.

These steering inequalities provide a new general means of witnessing entanglement and EPR steering. Given an entangled pair of $N$-dimensional quantum systems, tomographic reconstruction of the density matrix requires on the order of $N^{4}$ measurements, though it offers a complete description of the bipartite state. Violating our steering inequalities on the other hand requires only on the order of $N^{2}$ measurements, and in some cases only on the order of $N$ measurements for our sum/difference steering inequalities \eqref{pmsteerdv}.

Though our entropic steering inequalities are general, they are not superior to all other forms of steering inequality. As shown in Sec.~7 and Ref.~\cite{Cavalcanti2009}, for two-qubit systems, there are states which will violate a variance-based EPR-steering inequality and fail to violate the corresponding entropic 
steering inequality with the same set of measurements. The strength of our inequalities rests in their being expressed in terms of entropies, used in information theory. Other entropic inequalities have been used to show that EPR steering proves security in 1sDIQKD \cite{BranciardQKD2012}.

Our entropic steering inequalities also provide further evidence that there may exist states which exhibit steering in only one direction. Some states (i.e., those in either off-diagonal quadrant of Figures~2a and 2b), violate our entropic steering inequality in one direction, but not in the other, even when using an optimal set of measurement bases. This is an analogue, for the discrete case, of a phenomenon that has been shown to occur in continuous-variable systems by Midgley \emph{et al.} \cite{Midgley2010}. Neither of these results are definitive proof of the existence of exclusively one-way steerable states. There could be inequalities that witness one-way steering where our entropic inequalities fail to do so. A general proof would require a necessary and sufficient criterion for one-way and for two-way steering, but the inequalities presented here could provide a direction for further research.

We gratefully acknowledge insightful discussions with Gregory A. Howland, as well as support from DARPA DSO InPho Grant No. W911NF-10-1-0404. CJB acknowledges support from ARO W911NF-09-1-0385 and NSF PHY-1203931. SPW acknowledges funding support from the Future Emerging Technologies FET-Open Program, within the 7$^{th}$ Framework Programme of the European Commission, under Grant No. 255914, PHORBITECH and Brazilian agencies CNPq, CAPES, FAPERJ and the INCT-Informa\c{c}\~ao Qu\^antica. EGC acknowledges funding support from an ARC DECRA DE120100559.

\bibliography{EPRbib8}

\end{document}